\documentclass{article}
\usepackage{amssymb,amsmath,latexsym,amsfonts,graphicx,subfigure,color}

\input epsf.tex

\def\vep{\varepsilon}

\def\be{\begin{equation}}
\def\ee{\end{equation}}
\def\bea{\begin{eqnarray}}
\def\eea{\end{eqnarray}}
\def\beas{\begin{eqnarray*}}
\def\eeas{\end{eqnarray*}}

\newtheorem{prop}{Proposition}[section]
\newtheorem{ass}{Assumption}[section]
\newtheorem{thm}{Theorem}[section]

\newtheorem{rem}{Remark}[section]

\newtheorem{example}{Example}

\def\be{\begin{equation}}
  \def\eeq{\end{equation}}
\def\<{\langle}
\def\>{\rangle}

\def\vep{\varepsilon }

\def\bold(#1){\textbf{#1}}
\def\roma(#1){\textrm{#1}}
\def\und(#1){\underline{#1}}
\def\ove(#1){\overline{#1}}
\def\mbold(#1){\mathbb{#1}}

\def\ee{\'{e}}

\title{\bf \LARGE Reduced-order observer design for a robotic manipulator}
\author{Andrea Cristofaro and Alessandro De Luca% <-this % stops a space
\thanks{%$^\star$Corresponding author, \texttt{andrea.cristofaro@diag.uniroma1.it}\newline 
Authors are with the Department of Computer, Control and Management Engineering, Sapienza University of Rome, Via Ariosto 25, 00185 Rome, Italy.  \texttt{\{cristofaro,deluca\}@diag.uniroma1.it}
}}

\date{}
\begin{document}

\maketitle
%%%%%%%%%%%%%%%%%
\begin{abstract}
This paper investigates the design of reduced-order observers for robotic manipulators. Observer stability conditions are obtained based on a Lyapunov analysis and the proposed observer is enhanced with a hybrid scheme that may adjust the gains to cope with possible unbounded velocities of the robot joints. Thanks to such hybrid strategy, the observer works accurately both for robots driven by open-loop controllers and by output feedback controllers. Numerical simulations illustrate the efficacy of the reduced-order observer in several scenarios, including a comparison with the performances of a classical full-order observer.
\end{abstract}

\thispagestyle{empty}
\pagestyle{empty}
\section{Introduction}
Robotic manipulators are typically equipped with encoders mounted at the joints, which provide a direct measure of joint positions. On the other hand, most control tasks require also a velocity feedback in order to be executed. A rough knowledge of the velocity can be obtained by numerical differentiation, but the risk of occurring in large errors due to noise is high. Implementing state observers using the available position measurements is usually a better option.

The literature on observer design for nonlinear control systems is fairly large and several approaches have been explored, see for instance \cite{tsinias1989observer, raghavan1994observer, dalla2000design, aghannan2003intrinsic, karagiannis2008invariant, hammouri2010high} and the references therein. A considerable number of such studies are devoted to the specific problem of observer design for robotic manipulators. In this regard, it is worth mentioning the papers \cite{nicosia1989high, nicosia1990robot, de1991sliding, celani2006luenberger}
among several other works. A common feature of such observers is that their design parameters depend, in a way or another, on a bound on the maximum achievable velocity. Such design weakness is somewhat mitigated by the fact that the bound can be automatically enforced when the observer is used to implement feedback control laws, such as point-to-point regulation or trajectory tracking. Furthermore, available observers for robotic manipulators are typically full-order estimators, whereas the development of reduced-order schemes seems limited to a handful of papers, such as \cite{khelfi1996reduced} which however still suffers from the requirement of a known velocity bound. Reduced-order observers are well recognized for having faster convergence rates and lower computational burden, as the only variables to be estimated are the ones which are actually not measured.
In this paper we propose the design of a reduced-order observer derived from the structure of \cite{nicosia1990robot}. In addition, hinging on the powerful setup of hybrid systems \cite{goebel2009hybrid}, we introduce a control strategy to schedule the observer gain in order to adjust it according to the velocity of the joints and thus removing the need for a global bound. In light of this, the proposed observer performs efficiently even when the robot is controlled in open-loop and the velocity of the joints grows unbounded.
The paper is structured as follows. The manipulator setup is described in Section~\ref{sec:setup}, where the model and the main assumptions are introduced. Section~\ref{sec:observer} is devoted to present the observer design and the relative stability analysis, whereas Section~\ref{sec:hybrid} illustrates the enhancement of the observer with a switching scheme for the output injection gain. Simulations are reported in Section~\ref{sec:Sim}, where the efficiency of the reduced-order observer is showcased through several examples. Finally, concluding remarks are proposed in Section~\ref{sec:conclusions}.

\section{Robotic manipulator setup}\label{sec:setup}
Consider a robotic manipulator with $n\geq 1$ rigid joints, described by the dynamical model \cite{siciliano2010robotics}
\begin{equation}\label{eq:secondorder_dyn}
M(q)\ddot{q}+C(q,\dot{q})\dot{q}+F\dot{q}+g(q)=\tau
\end{equation}
where $q\in\mathbb{R}^n$ is the vector of joint positions and $\dot{q}\in\mathbb{R}^n$ is the vector of joint angular speeds. The inertia matrix $M(q)=M^T(q)$ is assumed to be positive definite for any $q\in\mathbb{R}^n$ and there exists positive definite constant matrices $M_1,M_2\in\mathbb{R}^{n\times n}$ with
\begin{equation}\label{eq:inertia}
M_1\preceq M(q)\preceq M_2\quad \forall q\in\mathbb{R}^n.
\end{equation}
The term $C(q,\dot{q})\dot{q}$ encodes Coriolis and centrifugal effects, $F$ is a dissipation matrix, $g(q)$ represents the gravity force vector and $\tau$ is the torque at the joints.\smallskip
\begin{ass}\label{ass:conditions}
{\it The following conditions are fulfilled:
\begin{itemize}
\item The angular position $q$ is known\smallskip
\item The matrix $C(q,\dot{q})$ is factorized in such a way that $\dot{M}(q)-2C(q,\dot{q})$ is skew symmetric\smallskip
\item The matrix $F$ is positive semi-definite, i.e. $F\succeq0$
\end{itemize}}
\end{ass}\smallskip
Some interesting and useful facts can be observed about the matrix $C(q,\dot{q})$. First, for any $y,u,w\in\mathbb{R}^n$, the following identity holds
\begin{equation}\label{eq:Coriolis1}
C(y,u)w=C(y,w)u.
\end{equation} 
Moreover, the desired factorization of $C(q,\dot{q})$ is such that a bounded and nonnegative function $0\leq c_0(q)\leq \bar{c}_0$ can be found with the property
\begin{equation}\label{eq:Coriolis2}
\|C(q,\dot{q})\|\leq c_0(q)\|\dot{q}\|.
\end{equation}

For the next developments, it is convenient to rewrite the manipulator second-order dynamics (\ref{eq:secondorder_dyn}) as a first order system with state variables $x_1=q,\ x_2=\dot{q}$ and output $y=x_1$:
\begin{equation}\label{eq:manipulator}
\begin{array}{rl}
\dot{x}_1&=x_2\\
M(x_1)\dot{x}_2&=-C(x_1,x_2)x_2-Fx_2-g(x_1)+\tau\\
y&=x_1
\end{array}
\end{equation}
Note that, the state of the system is now described by the vector $x=\mathrm{col}(x_1,x_2)\in\mathbb{R}^{2n}$.
\section{Reduced-order observer}\label{sec:observer}
The sub-state $x_1$ being available for direct measurements, the aim of this section is to design a reduced-order observer for system (\ref{eq:manipulator}) providing an asymptotic estimate for the sub-state $x_2$, the latter corresponding to the angular velocity of the robot joints.

Let us then consider a reduced-order observer with the following structure
\begin{equation}\label{eq:observer}
\begin{array}{rl}
M(y)\dot{z}&=-C(y,\hat{x}_2)\hat{x}_2-F\hat{x}_2-g(y)\\&-\displaystyle M(y)\frac{\partial k(y)}{\partial y}\hat{x}_2 +\tau\smallskip\\
\hat{x}_2&=z+k(y)
\end{array}
\end{equation}
where $k:\mathbb{R}^n\mapsto\mathbb{R}^n$ is a differentiable mapping to be specified. 
\begin{rem}
{\it In the case of a constant inertia matrix $M$, i.e., with $C(q,\dot{q})\equiv0$, the observer (\ref{eq:observer}) reads as a classical reduced-order observer for a second-order linear system.}
\end{rem}
The dynamics of the estimation error $\varepsilon=x_2-\hat{x}_2$, with $\hat{x}_2$ provided by (\ref{eq:observer}), is given by
$$
M(y)\dot\vep=-C(y,x_2)x_2+C(y,\hat{x}_2)\hat{x}_2-F\vep-M(y)\frac{\partial k(y)}{\partial y}\vep
$$
Adding and subtracting $C(y,\hat{x}_2)x_2$, and then using the identity (\ref{eq:Coriolis1}), the dynamics can be rearranged as
$$
\begin{array}{rl}
M(y)\dot\vep&=\displaystyle(-C(y,x_2)x_2+C(y,x_2)\hat{x}_2)\smallskip\\
&\displaystyle+(-C(y,\hat{x}_2)x_2+C(y,\hat{x}_2)\hat{x}_2)\smallskip\\
&\displaystyle-F\vep-M(y)\frac{\partial k(y)}{\partial y}\vep\smallskip\\
&=\displaystyle-C(y,x_2)\vep-C(y,\hat{x}_2)\vep-F\vep-M(y)\frac{\partial k(y)}{\partial y}\vep
\end{array}
$$
Consider now the time-varying Lyapunov function candidate $$V(\vep,t)=\frac12\vep^TM(y(t))\vep,$$ which is clearly positive definite due to the uniform bounds
$
\lambda_1\|\vep\|^2\leq V(\vep,t)\leq
\lambda_2\|\vep\|^2
$
provided by (\ref{eq:inertia}), where
\begin{equation}\label{eq:boundsM}
\lambda_1=\min_{y\in\mathbb{R}^n}\frac{\lambda_{\min}(M(y))}{2},\quad \lambda_2=\max_{y\in\mathbb{R}^n}\frac{\lambda_{\max}(M(y))}{2}
\end{equation}
Evaluating the derivative along the error system trajectory yields
$$
\begin{array}{rl}
\dot{V}(\vep,t)&=\displaystyle\frac12\vep^T\dot{M}(y)\vep+\vep^TM(y)\dot\vep\smallskip\\
&=\displaystyle\underbrace{\vep^T\left(\frac{\dot{M}(y)}2-C(y,x_2)\right)\vep}_{=0}-\vep^TC(y,\hat{x}_2)\vep\\&\displaystyle-\vep^T\left(F+M(y)\frac{\partial k(y)}{\partial y}\right)\vep
\end{array}
$$
In order to proceed with the stability analysis, we need to select a suitable gain function $k(y)$. Several options are available, as shown in the following subsections.
\subsection{Constant gain}
The simplest yet effective strategy is picking $k(y)$ as a linear function $k(y)=k_0y$ with $k_0>0$. This leads to the simplification 
\begin{equation}\label{eq:linear_gain}\frac{\partial k(y)}{\partial y}=k_0I_{n\times n}.\end{equation} Observing that $C(y,\hat{x}_2)=C(y,x_2-\vep)$, and that the bound $\|C(y,r)\|\leq c_0(y)\|r\|$ holds true, a bound on the Lyapunov function derivative can be obtained as follows
\begin{equation}\label{eq:Lyapunov_x2}
\begin{array}{rl}
\dot{V}&\!\!=\vep^T\left(-C(y,x_2)+C(y,\vep)-F-M(y)k_0\right)\vep\smallskip\\
&\!\!\!\leq \left(c_0(y)(\|x_2\|\!+\!\|\vep\|)-\lambda_{\min}(F)-k_0\lambda_{\min}(M(y))\right)\|\vep\|^2
\end{array}
\end{equation}
Next assumption is made to eliminate the dependency on $x_2$ in the right-hand side.\smallskip
\begin{ass}\label{ass:vmax}
{\it A known bound is available for the angular speed, i.e. $$\|\dot q\|=\|x_2\|\leq v_{\max}$$
for some known positive number $v_{\max}>0.$}
\end{ass}\smallskip
Thanks to such bound on the admissible angular velocity, we can infer the condition
\begin{equation}\label{eq:Lyapunov}
\dot{V}\leq\left(c_0(y)(v_{\max}\!+\!\|\vep\|)-\lambda_{min}(F)-k_0\lambda_{\min}(M(y))\right)\|\vep\|^2
\end{equation}
that enables for the selection of the gain $k_0.$ Pick arbitrarily $\eta>0$ and set
\begin{equation}\label{eq:k0}
k_0=\max_{y\in\mathbb{R}^n}\frac{c_0(y)(v_{\max}+\eta)-\lambda_{\min}(F)}{\lambda_{\min}(M(y))}
\end{equation}
Due to this choice, the right-hand side of (\ref{eq:Lyapunov}) is negative as long as the error $\vep$ is such that
$$
\|\vep\|<\eta\leq\frac{k_0\lambda_{\min}(M(y))+\lambda_{min}(F)-c_0(y)v_{\max}}{c_0(y)}
$$
Exploiting the bounds on the Lyapunov function, the error system turns out to be locally exponentially stable with region of attraction given by
\begin{equation}\label{eq:regionE}
\mathcal{E}=\left\{\vep\in\mathbb{R}^n:\|\vep\|<\eta\sqrt{\frac{\lambda_1}{\lambda_2}} \right\}
\end{equation}
Summarizing, the above derivations lead to the following statement.\smallskip
\begin{thm}
{\it Consider the manipulator system (\ref{eq:manipulator}), let Assumptions \ref{ass:conditions} and \ref{ass:vmax} hold, and let $\eta>0$ be fixed. Then the reduced-order observer (\ref{eq:observer}) with linear gain function $k(y)=k_0y$, where $k_0$ is assigned by (\ref{eq:k0}), provides a locally exponentially stable estimation error $\vep=x_2-\hat{x}_2$  with a region of attraction that contains the set $\mathcal{E}$ defined in (\ref{eq:regionE}).}
\end{thm}\smallskip
\begin{rem}
{\it It can be easily verified that, for any compact set $\mathcal{K}\subset \mathcal{E}$, the convergence rate $\varrho$ of the estimation error is given by
$$
\varrho=\eta-\sqrt{\lambda_2/\lambda_1}\max_{\vep\in\mathcal{K}}\|\vep\|
$$
}\end{rem}\smallskip
\begin{rem}
{\it A simpler, but more conservative, selection for the output injection gain $k_0$ can be done by setting 
$$
k_0=\frac{\bar{c}_0(v_{\max}+\eta)-\lambda_{\min}(F)}{2\lambda_1},
$$
the latter being obtained by taking, separately, the upper and lower bound for $c_0(y)$ and $\lambda_{\min}(M(y))$ respectively. It must be noticed, however, that selecting a large $k_0$ may lead to undesired effects due to noisy measurements and that there is in general a trade-off between stability properties and estimation performances in practice.}
\end{rem}\smallskip
\begin{rem}
{\it It is worth noticing that the conditions obtained for the reduced-order observer design, as well as the stability properties, are equivalent to those found for the full-order observer proposed by Nicosia-Tomei \cite{nicosia1990robot}.}
\end{rem}
\subsection{Nonlinear gain function}
Considering a nonlinear gain function $k(y)$, rather than a scalar and constant gain as in the case described earlier, might provide additional degrees of freedom in the observer design. For instance, it must be pointed out that the bound~(\ref{eq:k0}) is somewhat conservative and oversized as it is based on the least achievable eigenvalue for the inertia matrix. Furthermore, the spectrum of the inertia matrix can have large variations as the joint angles range over the admissible set, and the  eigenvalues typically achieve their maximum and minimum for singular configurations of the joints. As a consequence, in many scenarios, the observer is likely to provide an accurate estimation even if the gain is selected considerably lower than the bound given in (\ref{eq:k0}). To overcome this issue and design an estimator with a tighter output injection gain, it is possible to adopt more sophisticated strategy. We briefly present here two possible approaches to pursue, one devoted to limiting the deformation associated to the inertia matrix and the other one to seeking for a primitive of its inverse. Both strategies involve the formulation and the solution of mathematical problems which cannot be tackled by standard methods, and are worth a deeper investigation.
\begin{itemize}
\item {\it Deformation reduction}. The underlying idea of this strategy is to minimize the area deformation associated to the linear mapping $M(y)$, that is making the shape of level sets of the quadratic form $\vep^T [M(y)\frac{\partial k(y)}{\partial y}]\vep$ as close as possible to a sphere.
To this goal, we can select $k(y)=k_0K_1(y)$ with $k_0>0$ and $K_1(y)$ such that
$$ 
Q(y)=\left(M(y)\frac{\partial K_1(y)}{\partial y}\right)+\left(M(y)\frac{\partial K_1(y)}{\partial y}\right)^T>0
$$
and the ratio $$
\frac{\min_{y\in\mathbb{R}^n}{\lambda_{\min}(Q(y))}}
{\max_{y\in\mathbb{R}^n}{\lambda_{\max}(Q(y))}}$$
is maximized (ideally is equal to 1).\smallskip
\item {\it Matrix inversion}. The idea of the second strategy is based on the observation that, if $\frac{\partial k(y)}{\partial y}$ is chosen as the inverse of $M(y)$ multiplied by a constant gain, the resulting quadratic form can be shaped arbitrarily.  However, it is well known that a vector function corresponds to the gradient of a scalar function only under some rather conservative conditions. Nevertheless, we can attempt to approximate the inverse by selecting $k(y)=k_0K_1(y)$ with $k_0>0$ and $K_1(y)$ such that
$$
\frac{\partial K_1(y)}{\partial y}=M(y)^{-1}+Q(y)
$$
where $\|Q(y)\|<<1/\lambda_2$ and $\lambda_2$ is defined in (\ref{eq:boundsM}).
\end{itemize}
\section{A switching logic for gain scheduling}\label{sec:hybrid}
\begin{figure}[t!]
 \centering
 \includegraphics[width=0.9\columnwidth]{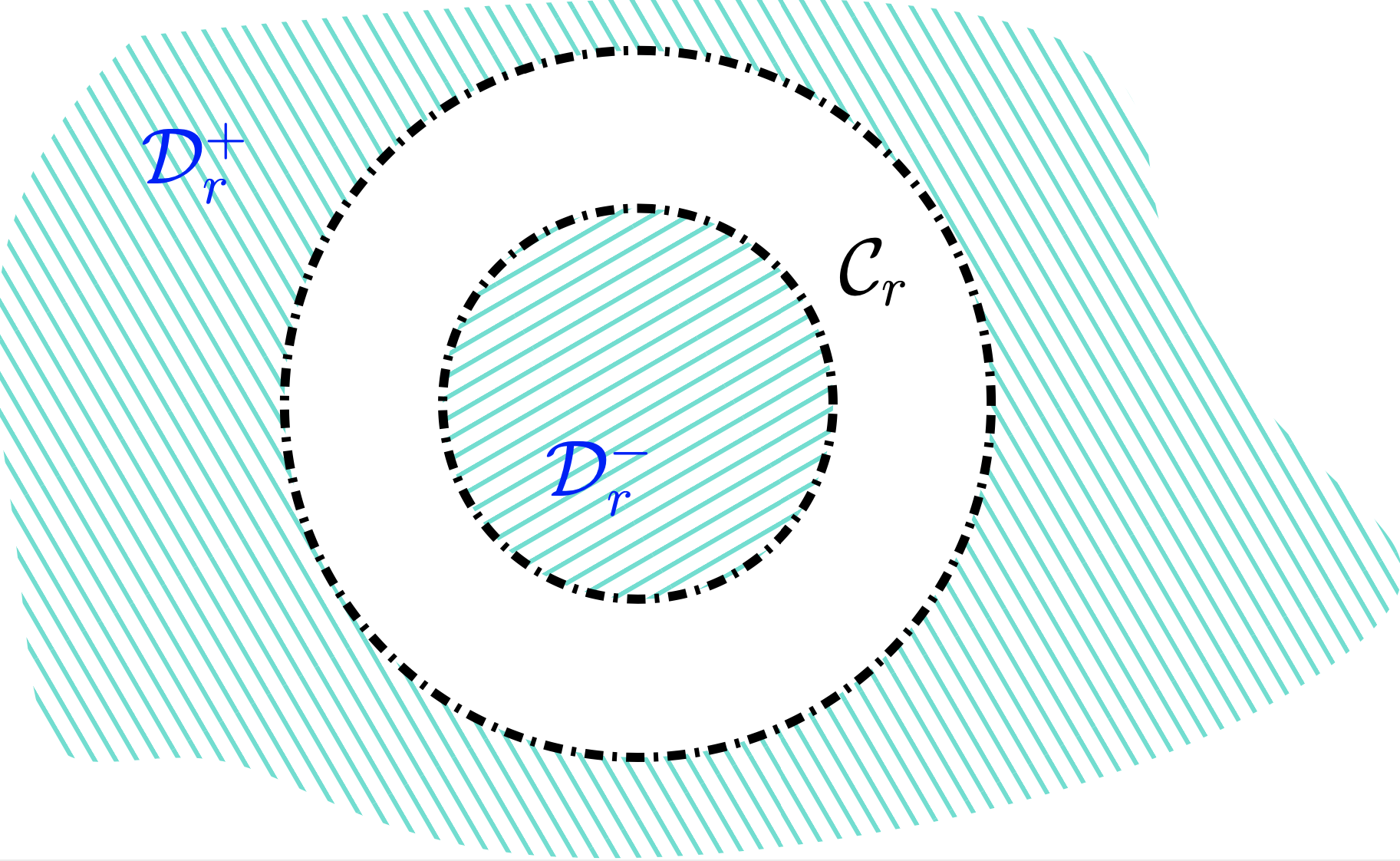}
 \caption{Illustration of flow and jump sets}\label{fig:hybrid}
 \end{figure}

As shown in \cite{nicosia1990robot}, when using the observer to implement an output feedback control with estimated velocities, the bound on the joint velocity can be automatically guaranteed. Conversely, when the observer is considered as a {\it standalone}, Assumption \ref{ass:vmax} is pivotal to guarantee negative definiteness of the Lyapunov function $V(\vep)$. In this section, a switching logic is proposed to bypass this issue and enable the observer to be locally asymptotically stable for arbitrary values of the angular velocity.\smallskip\\
Picking $\bar{v}>0$, let us consider the following decomposition of the velocity space $\mathbb{R}^n$:
$$
\begin{array}{c}
\bar{\mathcal{D}}_0=\{z\in\mathbb{R}^n:\|z\|\leq \bar{v}\}\smallskip\\
\bar{\mathcal{D}}_r=\{z\in\mathbb{R}^n:r\bar{v}<\|z\|\leq (r+1)\bar{v}\}\quad r=1,2,...
\end{array}
$$
Clearly we have $\bar{\mathcal{D}}_r\cap\bar{\mathcal{D}}_\ell=\emptyset$ for any $k\neq \ell$ and 
$$
\mathbb{R}^n=\bigcup_{r\in\mathbb{N}}\bar{\mathcal{D}}_r
$$
The idea is to increase/decrease the observer gain when the value of velocity $\dot x$ drifts from one region to another. Increasing the gain allows to keep the negative sign in the Lyapunov condition $\dot{V}(\vep)<0$, while a decrease limits the use of unnecessary efforts and avoids possible side effects such as noise amplification. However, two obstacles prevent the aforementioned strategy to be implemented. The first issue is that, without any hysteresis logic, when the velocity lies too close to the boundary between two regions the gain might repeatedly switch, this providing chattering  and deterioration of performances. The second and major problem is that we do not measure the velocity (the estimation of it being indeed our main goal), and thus we cannot have a direct knowledge of the index~$k$ describing the region $\bar{\mathcal{D}}_r$ where the true velocity actually lies. Nevertheless, a modified and feasible strategy can be successfully implemented by making use of the structure of the region of attraction $\mathcal{E}$ in (\ref{eq:regionE}) for the error system and applying a {\it bootstrap} argument. \\ \\
We can observe that for initial conditions $(x_2(0)-\hat{x}_2(0))\in\mathcal{E}$, as long as the Lyapunov function derivative $\dot{V}(\vep)$ is negative definite, the following estimate holds true
\begin{equation}\label{eq:bounds_x2}
\max\{0,\|\hat{x}_2\|-\eta\}\leq \|x_2\|\leq \|\hat{x}_2\|+\eta
\end{equation}
 We can use such lower and upper estimates to define suitable, and verifiable, switching conditions. With reference to Fig.~\ref{fig:hybrid}, pick $\bar{v}>2\eta$ and define the families of closed sets
 \begin{equation}\label{eq:jumpsets}
 \begin{array}{c}
 \mathcal{D}^{+}_r=\{z\in\mathbb{R}^n:\|z\|\geq r\bar{v}-\eta\} \smallskip \\
 \mathcal{D}^{-}_r=\{z\in\mathbb{R}^n:\|z\|\leq (r-1)\bar{v}+\eta\}
 \end{array}
 \end{equation}
 with $r\in\mathbb{N}.$ Accordingly, let us define $\mathcal{C}_r$ as the closure of the complementary set
 $$
 \mathcal{C}_r:=\overline{\mathbb{R}^n\setminus(\mathcal{D}^{+}_r\cup\mathcal{D}^{-}_r)}.
 $$
 
 Based on the previous decomposition and similarly to the principles used, for example, to implement scheduled anti-windup policies \cite{cristofaro2019switched}, we can introduce a logic state variable $r\in\mathbb{N}$, define the observer with a scheduled gain 
 \begin{equation}\label{eq:kr}
 k_r=\max_{y\in\mathbb{R}^n}\frac{c_0(y)(\eta+r\bar{v})-\lambda_{\min}(F)}{\lambda_{\min}(M(y))}
\end{equation}
 and let $r$ evolve according to the hybrid dynamics
\begin{equation}\label{eq:hybrid_dyn}
 \begin{array}{ll}
 \dot{r}=0& \hat{x}_2\in\mathcal{C}_r\\
 r^+=r+1&\hat{x}_2\in\mathcal{D}^+_r\\
  r^+=r-1&\hat{x}_2\in\mathcal{D}^-_r
 \end{array}
\end{equation}
 Following the classical notation used in the context of hybrid systems \cite{goebel2009hybrid}, the set 
 $$\bigcup_{r\in\mathbb{N}}\mathcal{C}^r\times\{r\}$$ is referred to as the {\it flow set}, whereas $$\bigcup_{r\in\mathbb{N}}(\mathcal{D}^{+}_r\cup\mathcal{D}^{-}_r)\times\{r\}$$ is the {\it jump set}. %(see Fig. \ref{fig:hybrid} for a graphical illustration).
\smallskip\\
 Suppose now $\vep(0)\in\mathcal{E}$ so that, in particular, $\|\vep(0)\|\leq\eta$ and the bounds (\ref{eq:bounds_x2}) hold. In addition, without loss of generality, let us assume\footnote{If this is not the case, the hybrid dynamics is such that the logic state $r$ immediately increases, if needed, up to a value large enough to guarantee that the gain $k_r$ defined in (\ref{eq:kr}) is able to cope with the size of the joint velocities.} that $r(0)$ satisfies $\|x_2(0)\|\leq r(0)\bar{v}$. The bootstrap argument then applies: the derivative of the Lyapunov function $\dot{V}(\vep(0))$ in~(\ref{eq:Lyapunov_x2}) is negative definite, the estimation error $\vep(t)$ remains confined in a ball of radius $\eta$ and this, in turn, implies that estimates~(\ref{eq:bounds_x2}) keep holding and that the sets~(\ref{eq:jumpsets}) are well-defined for any $t>0.$ In other words, a correct initialization of the observer~(\ref{eq:observer}) with gain $k(y)=k_ry$ enhanced by the hybrid dynamics (\ref{eq:kr})-(\ref{eq:hybrid_dyn}) is enough to guarantee that the error system is locally asymptotically stable independently of the bound on the joint velocity. Let us summarize the above discussion in the following statement.
 \begin{prop}{\it
 Consider the manipulator system (\ref{eq:manipulator}) and let Assumption \ref{ass:conditions} hold. Let the velocity $x_2(t)$ be such that $$
 \lim\sup_{t\rightarrow\infty} \|x_2(t)\|\leq \tilde{v},
 $$
 for some constant, yet unknown, number $\tilde{v}>0$,. Consider the reduced-order observer (\ref{eq:observer}) with gain $k(y)=k_ry$ where $k_r$ is defined in (\ref{eq:kr}) and the logic state $r$ is governed by the hybrid dynamics (\ref{eq:hybrid_dyn}). Then the dynamics of the estimation error $\vep=x_2-\hat{x}_2$ is locally asymptotically stable with a region of attraction containing the set $\mathcal{E}$ defined in (\ref{eq:regionE}) . Furthermore, there exists a sufficiently large natural number $N_{\tilde{v}}\in\mathbb{N}$ such that the logic state is ultimately bounded with 
 $$
 \lim\sup_{t\rightarrow\infty}r(t)\leq N_{\tilde{v}}.
 $$}
 \end{prop}
 \section{Simulations}\label{sec:Sim}
 To illustrate the performance of the reduced observer, we consider the model of a two-link robot arm, described by the following physical parameters\footnote{The links are supposed to satisfy a thin rod assumption, with barycenter in $d=L/2$ and barycentric inertia expressed by $I=(1/12)mL^2$.}
 \begin{table}[htp]
\begin{center}
\begin{tabular}{|c|c|}
\hline
mass link 1 & 10\ Kg\\
\hline
mass link 2 & 20\ Kg\\
\hline
length link 1 &1\ m\\
\hline
length link 2 & 1.5\ m\\
\hline
damping link 1 & 0.1\ Kg/s\\
\hline
damping link 2 & 0.3\ Kg/s\\
\hline
\end{tabular}
\end{center}
\caption{Physical parameters of the two-link robot arm}
\label{default}
\end{table}%

\begin{figure}[h!]
 \centering
 \includegraphics[width=0.9\columnwidth]{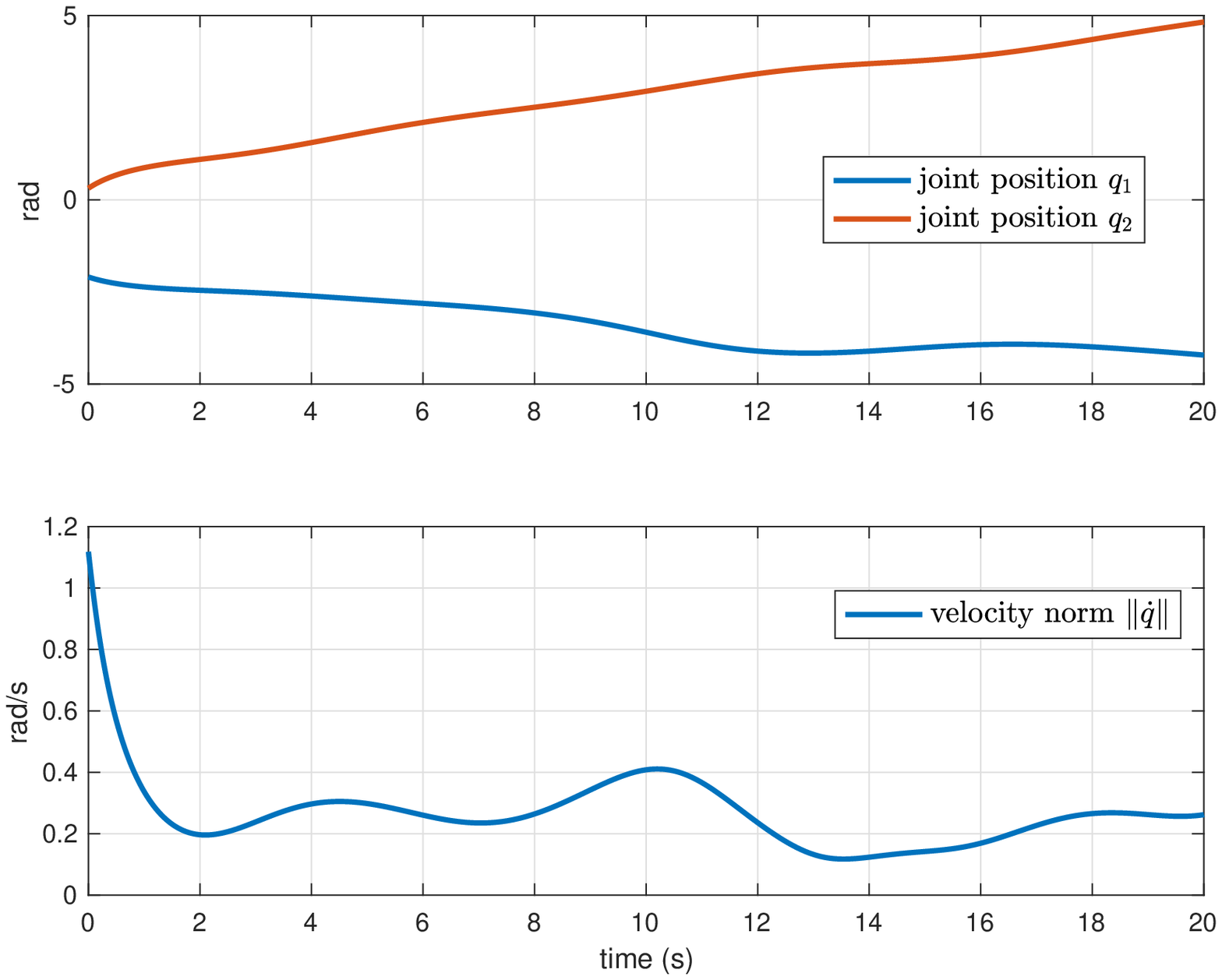}
 \caption{Example \ref{ex:open1}: Joint positions (a) and velocity norm (b)}\label{fig:open_pos}
 \end{figure}

\begin{figure}[h!]
 \centering
 \includegraphics[width=0.9\columnwidth]{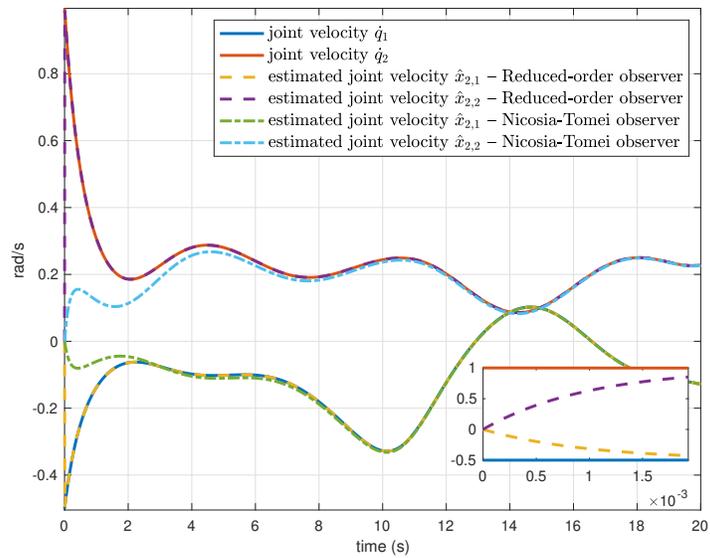}
 \caption{Example \ref{ex:open1}: Actual vs estimated angular velocity of joints}\label{fig:open_vel}
 \end{figure}

The initial conditions $x_1(0)=[q_1(0)\ q_2(0)]^\top,\ x_2(0)=[\dot{q}_1(0)\ \dot{q}_2(0)]^\top$ for the robot are assumed as
$$
\begin{array}{c}
q_1(0)=-\frac{2\pi}{3}\ \mathrm{rad},\quad q_2(0)=\frac{\pi}{10}\ \mathrm{rad},\smallskip\\ \dot {q}_1(0)=-0.5\ \mathrm{rad/s},\quad \dot{q}_2(0)=1\ \mathrm{rad/s}
\end{array}
$$
and the observer has been initialized at zero for the sake of simplicity, i.e., $\hat{x}_2(0)=[0\ 0]^\top\ \mathrm{rad/s}$. Three scenarios have been considered, addressing both the case of open-loop controllers and the regulation problem via dynamic output feedback.
\begin{example}{\it Open-loop control with bounded velocity.}\label{ex:open1} Let us assume the robot motion to be driven by the open loop control law
$$
\tau(t)=g(q)+\begin{bmatrix}
\cos(t/2)\\
-\cos(t)
\end{bmatrix}
$$
which includes a term compensating for the gravity.
Using the linear formulation \eqref{eq:linear_gain}, the observer parameters have been chosen as follows
$$
\eta=1\ \mathrm{rad/s},\ v_{\max}=1.5\ \mathrm{rad/s},
$$
with the constant gain $k_0$ given by \eqref{eq:k0}, this providing a stability region that contains the ball of radius~1 subject to the initial guess $|x_2(0)|\leq1.5\ \mathrm{rad/s}$. In this regard let us point out that, due to the particular choice of control torques, the velocity remains within the bound during the whole evolution of the system (see  Figure~\ref{fig:open_pos}(b)). To better highlight the performances of the reduced-order, a comparison with the results obtained using the observer by Nicosia-Tomei  \cite{nicosia1990robot} is reported. The latter has been implemented with the same gain $k_0$ used for the reduced-order observer, as the choice of such parameter determines the range of admissible velocities for the stability of the observer and dictates the convergence rate of the error dynamics.
%For the sake of simplicity and without loss of generality, the robot motion has been supposed to be driven by the open loop control law
%$$
%\tau(t)=g+\begin{bmatrix}
%\sin(t)\\
%1+\sin(2t)
%\end{bmatrix}
%$$
The simulation results are reported in Figures~\ref{fig:open_pos}-\ref{fig:open_vel}. In particular, Figure~\ref{fig:open_pos}(a) depicts the evolution of the position of the two joints under the considered control law, while the time-history of actual and estimated velocities is shown in Figure \ref{fig:open_vel}. Although both observers converge to the actual velocity of the joints as expected, it is clearly visible that the reduced-order observer is much faster than the full-order observer by \cite{nicosia1990robot}. This observation suggests that, under the same stability conditions and with a comparable region of attraction, the reduced-order observer performs better in terms of convergence rate.
\end{example}

\begin{figure}[h!]
 \centering
 \includegraphics[width=0.9\columnwidth]{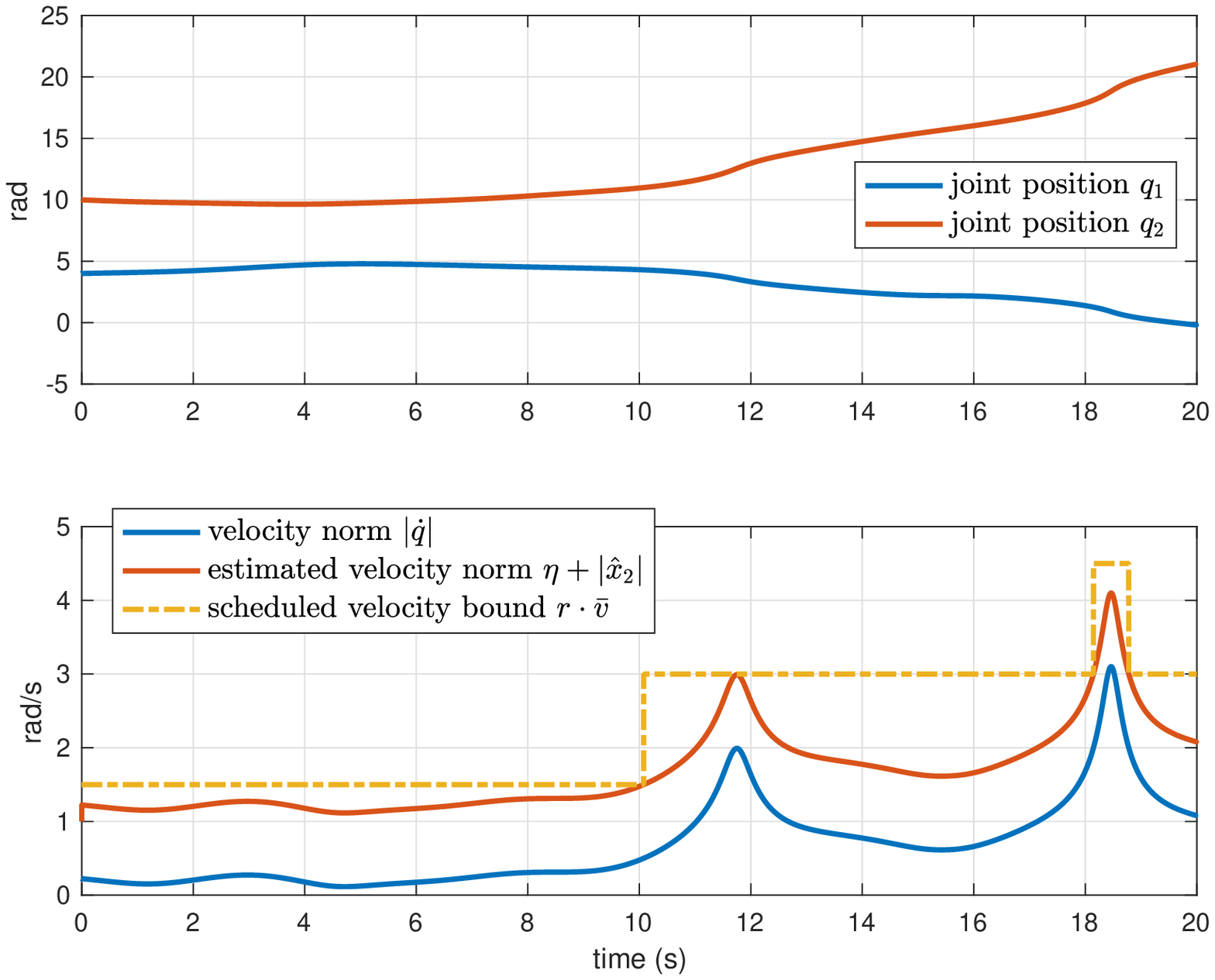}
 \caption{Example \ref{ex:open2}: Joint positions (a) and velocity norm (b)}\label{fig:hyb_pos}
 \end{figure}

\begin{figure}[h!]
 \centering
 \includegraphics[width=0.9\columnwidth]{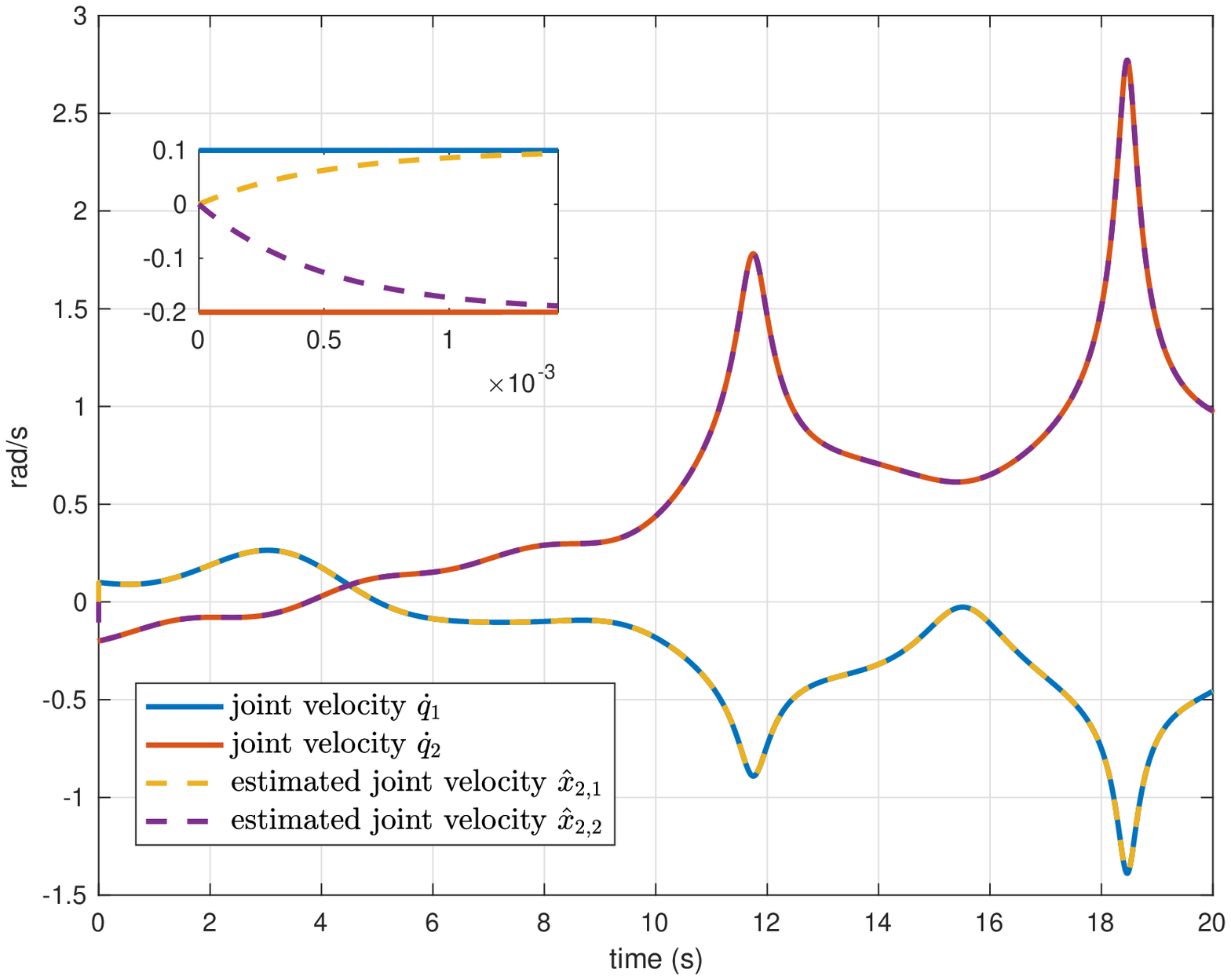}
 \caption{Example \ref{ex:open2}: Actual vs estimated velocity of joints}\label{fig:hyb_vel}
 \end{figure}

\begin{figure}[h!]
 \centering
 \includegraphics[width=0.9\columnwidth]{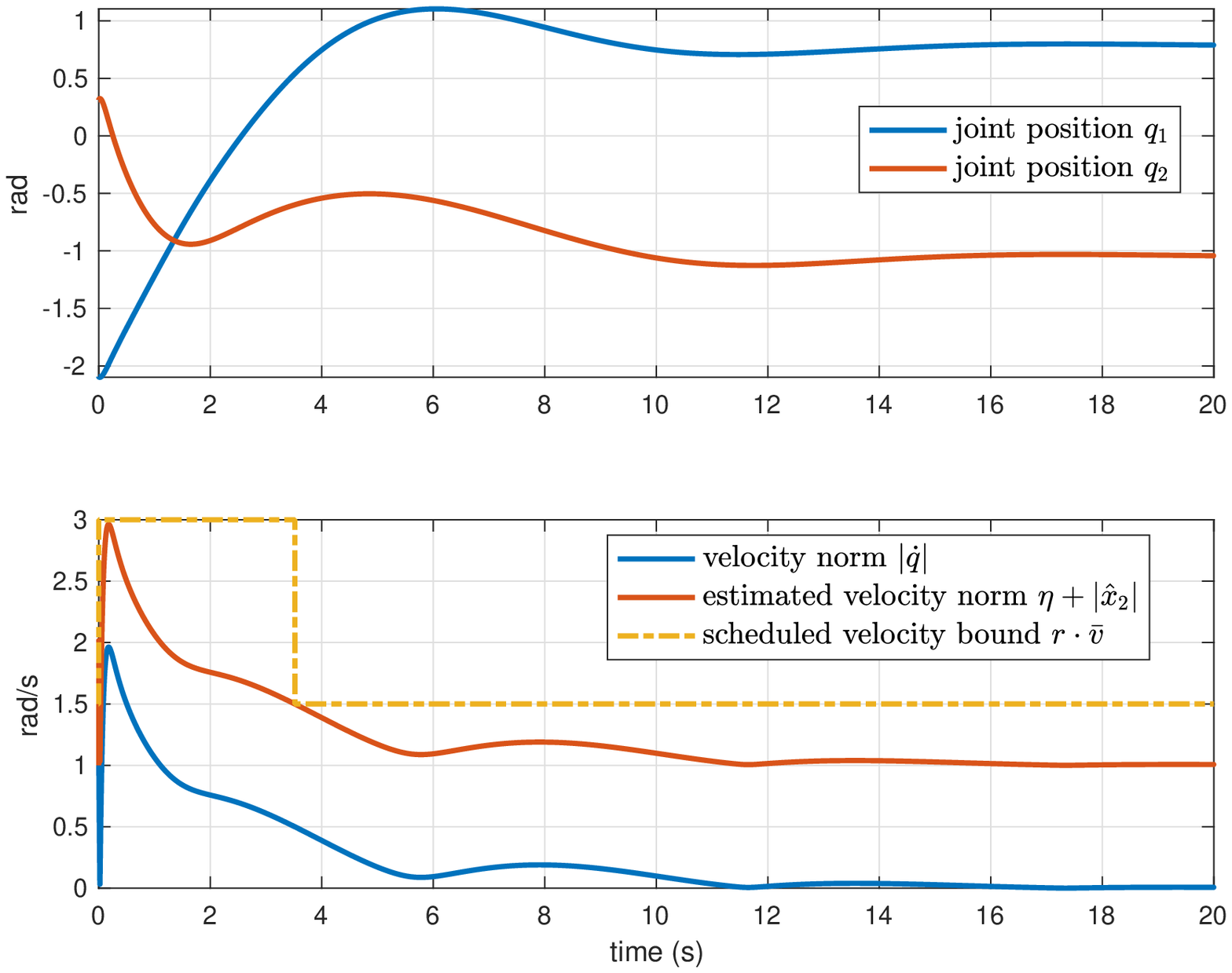}
 \caption{Example \ref{ex:PD}: Joint positions (a) and velocity norm (b)}\label{fig:PD_pos}
 \end{figure}

\begin{figure}[h!]
 \centering
 \includegraphics[width=0.9\columnwidth]{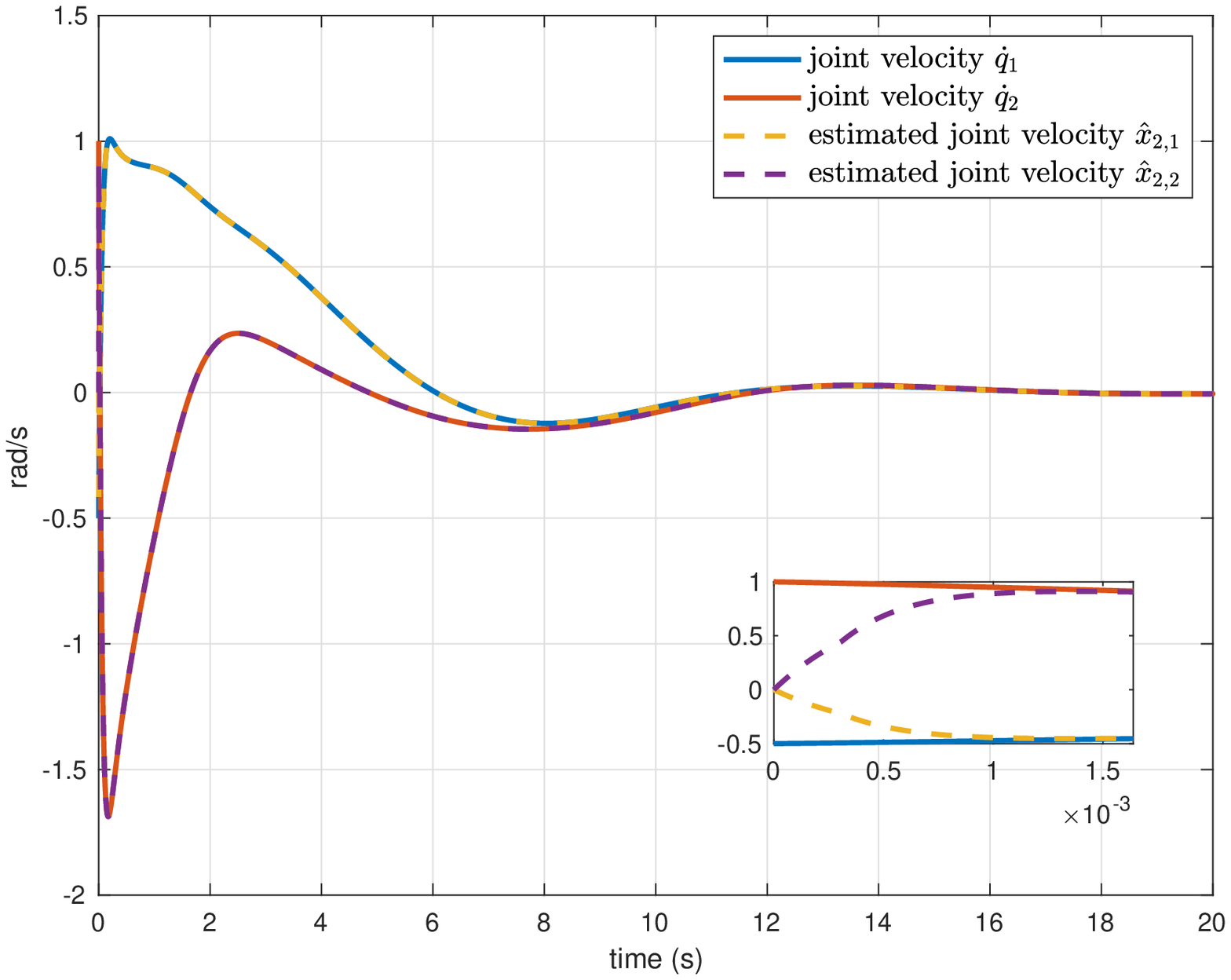}
 \caption{Example \ref{ex:PD}: Actual vs estimated velocity of joints}\label{fig:PD_vel}
 \end{figure}

\begin{example}{\it Open-loop control with unbounded velocity.}\label{ex:open2} Let us consider now another open-loop control given by
$$
\tau(t)=g(q)+\begin{bmatrix}
\sin(t)\\
1+\sin(2t)
\end{bmatrix}
$$
Unlike in the previous case, under such control input the velocity grows unbounded and therefore we need to resort to the hybrid scheme with the scheduled gain $k_r$ proposed in \eqref{eq:kr}-\eqref{eq:hybrid_dyn} in order to guarantee the stability of the observer. The velocity bound has been fixed ad $\bar{v}=1.5$ to be consistent with the setup of the first example and, accordingly, the initial guess $r(0)=1$ has been made for the logic state~$r$. Figure~\ref{fig:hyb_pos}(a) reports the behaviour of the position of the joints and
Figure~\ref{fig:hyb_vel} clearly shows the efficiency of the proposed estimation technique: after a very short transient (better highlighted in the top-left box), the observer states perfectly match the velocities of the two joints. Finally, the plots of the velocity norm and the bounds used to compute the scheduled observer gain are given in Figure~\ref{fig:hyb_pos}(b): as expected, thanks to a successful initial guess, the norm of the actual velocity remains bounded by the estimated velocity norm, and this allows to correctly trigger the jumps of the logic state $r$ and update the observer gain accordingly.
\end{example}
\begin{example}{\it Dynamic output feedback.}\label{ex:PD}
Let us consider now a different scenario, where the robot is controlled by a PD feedback law for the regulation of the joint positions to given setpoints. Since the robot velocity is not directly measured, the velocity feedback has been implemented using the estimation provided by the observer. In particular the following control law has been fed to the system
$$
\tau(t)=g(q)+K_{\mathrm{p}}(x_{\mathrm{ref}}-q(t))-K_{\mathrm{d}}\hat{x}_2(t)
$$
with reference setpoint $x_{\mathrm{ref}}=[\pi/4\ -\pi/3]^\top$ and gain matrices
$
K_{\mathrm{p}}=\mathrm{diag}(40,20),$ $ K_{\mathrm{d}}=\mathrm{diag}(60,30).
$ 

Keeping the observer parameters
$\eta=1\ \mathrm{rad/s}$ and $\bar{v}=1.5\ \mathrm{rad/s}$, 
 the logic state $r$ is initialized by setting $r(0)=1$. The simulation results are reported in Figures \ref{fig:PD_pos}-\ref{fig:PD_vel}. The evolution of the position of the two joints is shown in Figure~\ref{fig:PD_pos}(a), where the convergence towards the desired values can be noticed. The estimation performance of the reduced-order observer is well illustrated in Figure~\ref{fig:PD_vel}: despite the initial peak of the angular velocity of joints, the observer is able to perfectly reconstruct the state of the system after a short transient. This can be also appreciated in Figure~\ref{fig:PD_pos}(b), showing that the bounds on the velocity are computed successfully and that the switching of the observer gain is activated consistently. We notice that, after the initial transient, the velocity $\dot{q}$ converge to zero and, accordingly, the logic state $r$ jumps down to the least achievable value.

\end{example}
% \begin{figure}[h!]
% \centering
% \includegraphics[width=\columnwidth]{open1_position.eps}
% \caption{Example \ref{ex:open1}: Angular position of joints}\label{fig:open1_pos}
% \end{figure}
%  \begin{figure}[h!]
% \centering
% \includegraphics[width=\columnwidth]{open1_velocity.eps}
% \caption{Example \ref{ex:open1}: Actual vs estimated angular velocity of joints}\label{fig:open1_vel}
% \end{figure}
% \begin{figure}[h!]
% \centering
% \includegraphics[width=\columnwidth]{hyb_position.eps}
% \caption{Example \ref{ex:open2}: Angular position of joints}\label{fig:position_hyb}
% \end{figure}
%   \begin{figure}[h!]
% \centering
% \includegraphics[width=\columnwidth]{hyb_velocity.eps}
% \caption{Example \ref{ex:open2}: Actual vs estimated velocity of joints}\label{fig:velocity_hyb}
% \end{figure}
%   \begin{figure}[h!]
% \centering
% \includegraphics[width=\columnwidth]{hyb_bounds.eps}
% \caption{Example \ref{ex:open1}: Velocity bounds}\label{fig:bounds_hyb}
% \end{figure}
%
%  \begin{figure}[h!]
% \centering
% \includegraphics[width=\columnwidth]{PD_position.eps}
% \caption{Example \ref{ex:PD}: Angular position of joints}\label{fig:position_PD}
% \end{figure}
%   \begin{figure}[h!]
% \centering
% \includegraphics[width=\columnwidth]{PD_velocity.eps}
% \caption{Example \ref{ex:PD}: Actual vs estimated velocity of joints}\label{fig:velocity_PD}
% \end{figure}
%   \begin{figure}[h!]
% \centering
% \includegraphics[width=\columnwidth]{PD_bounds.eps}
% \caption{Example \ref{ex:PD}: Velocity bounds}\label{fig:bounds_PD}
% \end{figure}

\section{Conclusions and discussion}\label{sec:conclusions}
The problem of velocity observer design for robotic manipulators has been addressed in this paper. Taking inspiration from the classical full-order observer \cite{nicosia1990robot}, we developed a reduced-order estimator with the aim of lowering the computational burden and increasing the convergence rate. Furthermore, thanks to the coupling with a hybrid scheme for adjusting the observer gains, it has been possible to avoid the requirement of a global bound on the robot velocity, which is indeed a typical design limitation for this class of mechanical systems. Such improvement enables for the application of the observer also in open-loop control regimes that are usually not able to guarantee bounded velocities. Numerical simulations support and validate the theoretical findings by the illustration of the observer behaviour in several scenarios.
Furthermore, the use of the reduced-order observer to implement momentum-based filters \cite{de2005sensorless} for detection purposes is object of current research.

 \bibliography{reduced}

\begin{thebibliography}{10}
\providecommand{\url}[1]{#1}
\csname url@rmstyle\endcsname
\providecommand{\newblock}{\relax}
\providecommand{\bibinfo}[2]{#2}
\providecommand\BIBentrySTDinterwordspacing{\spaceskip=0pt\relax}
\providecommand\BIBentryALTinterwordstretchfactor{4}
\providecommand\BIBentryALTinterwordspacing{\spaceskip=\fontdimen2\font plus
\BIBentryALTinterwordstretchfactor\fontdimen3\font minus
  \fontdimen4\font\relax}
\providecommand\BIBforeignlanguage[2]{{%
\expandafter\ifx\csname l@#1\endcsname\relax
\typeout{** WARNING: IEEEtran.bst: No hyphenation pattern has been}%
\typeout{** loaded for the language `#1'. Using the pattern for}%
\typeout{** the default language instead.}%
\else
\language=\csname l@#1\endcsname
\fi
#2}}

\bibitem{tsinias1989observer}
J.~Tsinias, ``Observer design for nonlinear systems,'' \emph{Systems \& Control
  Letters}, vol.~13, no.~2, pp. 135--142, 1989.

\bibitem{raghavan1994observer}
S.~Raghavan and J.~K. Hedrick, ``Observer design for a class of nonlinear
  systems,'' \emph{Int. J. of Control}, vol.~59, no.~2, pp. 515--528, 1994.

\bibitem{dalla2000design}
M.~Dalla~Mora, A.~Germani, and C.~Manes, ``Design of state observers from a
  drift-observability property,'' \emph{IEEE Trans. on Automatic Control},
  vol.~45, no.~8, pp. 1536--1540, 2000.

\bibitem{aghannan2003intrinsic}
N.~Aghannan and P.~Rouchon, ``An intrinsic observer for a class of lagrangian
  systems,'' \emph{IEEE Trans. on Automatic Control}, vol.~48, no.~6, pp.
  936--945, 2003.

\bibitem{karagiannis2008invariant}
D.~Karagiannis, D.~Carnevale, and A.~Astolfi, ``Invariant manifold based
  reduced-order observer design for nonlinear systems,'' \emph{IEEE Trans. on
  Automatic Control}, vol.~53, no.~11, pp. 2602--2614, 2008.

\bibitem{hammouri2010high}
H.~Hammouri, G.~Bornard, and K.~Busawon, ``High gain observer for structured
  multi-output nonlinear systems,'' \emph{IEEE Trans. on automatic control},
  vol.~55, no.~4, pp. 987--992, 2010.

\bibitem{nicosia1989high}
S.~Nicosia and A.~Tornamb{\`e}, ``High-gain observers in the state and
  parameter estimation of robots having elastic joints,'' \emph{Systems \&
  control letters}, vol.~13, no.~4, pp. 331--337, 1989.

\bibitem{nicosia1990robot}
S.~Nicosia and P.~Tomei, ``Robot control by using only joint position
  measurements,'' \emph{IEEE Trans. on Automatic Control}, vol.~35, no.~9, pp.
  1058--1061, 1990.

\bibitem{de1991sliding}
C.~C. De~Wit and J.-J. Slotine, ``Sliding observers for robot manipulators,''
  \emph{Automatica}, vol.~27, no.~5, pp. 859--864, 1991.

\bibitem{celani2006luenberger}
F.~Celani, ``A {L}uenberger-style observer for robot manipulators with position
  measurements,'' in \emph{2006 14th Med. Conf. on Control and
  Automation}.\hskip 1em plus 0.5em minus 0.4em\relax IEEE, 2006, pp. 1--6.

\bibitem{khelfi1996reduced}
M.~F. Khelfi, M.~Zasadzinski, H.~Rafaralahy, E.~Richard, and M.~Darouach,
  ``Reduced-order observer-based point-to-point and trajectory controllers for
  robot manipulators,'' \emph{Control Engineering Practice}, vol.~4, no.~7, pp.
  991--1000, 1996.

\bibitem{goebel2009hybrid}
R.~Goebel, R.~G. Sanfelice, and A.~R. Teel, \emph{Hybrid dynamical
  systems}.\hskip 1em plus 0.5em minus 0.4em\relax Princeton University Press,
  2012.

\bibitem{siciliano2010robotics}
B.~Siciliano, L.~Sciavicco, L.~Villani, and G.~Oriolo, \emph{Robotics:
  modelling, planning and control}.\hskip 1em plus 0.5em minus 0.4em\relax
  Springer, 2010.

\bibitem{cristofaro2019switched}
A.~Cristofaro, S.~Galeani, S.~Onori, and L.~Zaccarian, ``A switched and
  scheduled design for model recovery anti-windup of linear plants,''
  \emph{European J. of Control}, vol.~46, pp. 23--35, 2019.

\bibitem{de2005sensorless}
A.~De~Luca and R.~Mattone, ``Sensorless robot collision detection and hybrid
  force/motion control,'' in \emph{Proc. of the 2005 IEEE international conf.
  on robotics and automation}.\hskip 1em plus 0.5em minus 0.4em\relax IEEE,
  2005, pp. 999--1004.

\end{thebibliography}
 \end{document}